\documentstyle[prd,epsfig,aps,preprint]{revtex}
\begin{document}

\draft

\title{\bf Transport coefficients and nonextensive statistics}
\author {J. R. Bezerra\footnote{zero@dfte.ufrn.br}, R. Silva\footnote
{raimundo@dfte.ufrn.br}, J. A. S. Lima\footnote
{limajas@dfte.ufrn.br}}

\smallskip
\address{~\\Universidade
Federal do Rio Grande do Norte,
\\Departamento de F\'{\i}sica,
Caixa Postal 1641, \\59072-970 Natal, RN, Brazil}

\date{\today}
\maketitle

\vskip 1.5cm
\begin{abstract}
We discuss the basic transport phenomena in gases and plasmas
obeying the $q$-nonextensive velocity distribution (power-law).
Analytical expressions for the thermal conductivity ($K_q$) and
viscosity ($\eta_q$) are derived by solving the Boltzmann equation
in the relaxation-time approximation. The available experimental
results to the ratio {$K_q$}/$\eta_q$ constrains the $q$-parameter
on the interval $0.74 \leq q \leq 1$. In the extensive limiting
case, the standard transport coefficients based on the local
Gaussian distribution are recovered, and due to a surprising
cancellation, the electric conductivity of a neutral plasma is not
modified.
\end{abstract}

\pacs{05.45.+b; 05.20.-y; 05.90.+m }

\newpage
\section{Introduction}

In the last few years, an increasing  attention has been paid to
possible nonextensive effects in the fields of thermodynamics and
statistical mechanics. The main motivation is the lack of a
comprehensive treatment including gravitational and Coulombian
fields for which the assumed additivity of the entropy present in
the standard approach is not valid [1-2]. Inspired on such
problems, as well as in the traditional ensemble theory,
Tsallis\cite{T88} proposed a remarkable $q$-parameterized
nonextensive entropic expression which reduces to the extensive
Gibbs-Jaynes-Shannon entropy in the limiting case $q=1$ (see
\cite{C99} for a regularly updated bibliography on this subject).

Later on, the first attempts exploring the kinetic route
associated to this nonextensive approach appeared in the
literature[5-10]. The original kinetic derivation advanced by
Maxwell\cite{M1860} was generalized to include power law
distributions as required by this enlarged framework. In
particular, it was shown that the equilibrium  velocity
$q$-distribution
\begin{equation}
\label{eq0} f_0(v) = B_q [1-(1-q){m v^2\over 2 k_B T}]^{1/(1-q)},
\end{equation}
is uniquely determined from two simple requirements\cite{SPL98}:
(i) isotropy of the velocity space, and (ii) a suitable
nonextensive generalization of the Maxwell factorizability
condition, or equivalently, the assumption that
$F(v)=f(v_x)f(v_y)f(v_z)$. The quantity $B_q$ above denotes the
$q$-dependent normalization constant whose expression on the
interval  $1/3<q \leq 1$ is
\begin{equation}\label{eq13}
B_q = n(1-q)^{1/2}\frac{5-3q}{2}\frac{3-q}{2}\frac{\Gamma
(\frac{1}{2}+{1\over 1-q})}{\Gamma({1\over
1-q})}\left(\frac{m}{2\pi k_BT}\right)^{3/2},
\end{equation}
where $n$ is the particle number density, $m$ is the mass and $T$
is the temperature. As expected, in the limiting case $q=1$, such
expressions reduce to the standard Maxwellian ones\cite{SPL98}.

More recently, the kinetic foundations of the above distribution
were investigated in a deeper level through the generalized
Boltzmann's transport equation\cite{LSP01}
\begin{equation}
\label{Beq} {\partial f\over\partial t} + {\bf v}\cdot{\partial
f\over\partial {\bf r}}+{{\bf F}\over m}\cdot{\partial f\over
\partial{\bf v}}=C_q(f),
\end{equation}
where $C_q$ is the $q$-nonextensive integral source term measuring
the change in $f$ due to collisions. This Boltzmannian like
approach incorporated the nonextensive effects using two different
ingredients. First, a new functional form to the kinetic local gas
entropy, and, second, a nonfactorizable distribution function for
the colliding pairs of particles whose physical meaning is quite
clear: the Boltzmann chaos molecular hypothesis is not valid in
this extended framework. It was also shown that the kinetic
version of the Tsallis entropy satisfies an $H_q$-theorem, and
more important still, the $q$-parameterized class of power law
velocity distributions emerged as the unique nonextensive solution
describing the equilibrium states.

In this paper we go one step further by computing the basic
nonextensive transport coefficients for dilute gases and plasmas.
Due to the intricate form of the new collisional term, we adopt
here the relaxation-time model proposed by Bhatnager, Gross and
Krook\cite{bgk} because it permits an exact mathematical
treatment. In spite of its limitation, the relaxation-time model
(henceforth BGK model) has been proved quite useful from a
methodological viewpoint because it yields the correct answer to
the problem in a first approximation, and as such, it can guide us
to the correct nonextensive results which must rigorously be
obtained using the full $q$-transport equation. As we shall see,
analytical expressions for the transport coefficients of a dilute
gas are readily obtained with basis on the local $q$-nonextensive
distribution. In particular, we show that the available
experimental results to the ratio {$K_q$}/$\eta_q$ constrains the
$q$-parameter on the interval $0.74 \leq q \leq 1$. As expected,
in the extensive limiting case, the classical expressions of the
transport coefficients are readily recovered. However, due to a
surprising cancellation, the standard electric conductivity of a
dilute neutral plasma is not modified. The analysis presented here
is also compared with independent calculations based on rather
different approaches\cite{boghosian,potiguar}.

\section{Nonextensive Transport Coefficients}

Let us now consider a dilute gas whose particles are acted by
external forces. In the BGK approximation the $q$-transport
equation (\ref{Beq}) reads :
\begin{equation}\label{eq1}
\frac{\partial{\it f}}{\partial{\it t}} + {\bf
v}\cdot\frac{\partial{\it f}}{\partial{\bf r}} + \frac{\bf
F}{m}\cdot\frac{\partial{\it f}}{\partial{\bf v}} =
-\frac{f-f_0}{\tau},
\end{equation}
where $\bf{F}$ is the external force and $m$ is the mass of the
particles. In this approximation,  $f_0$ is the local equilibrium
distribution function and $\tau$ is a relaxation time which is a
number of the order of the collision time. We recall that under
stationary conditions the distribution function does not depend on
time, however, the  concentration, and the temperature, are local
quantities. The whole system is assumed to be out but close to the
local stationary equilibrium state\cite{GM}. In particular, this
means that the resulting nonequilibrium  distribution, $f({\bf
r},{\bf v})$, is only slightly different from the equilibrium
unperturbed stationary distribution, $f_0({\bf r},{\bf v})$, and
can be approximated as
\begin{equation}\label{oi}
f({\bf r},{\bf v})=f_0({\bf v},{\bf r})+g({\bf v},{\bf r}),\quad
|g|\ll f_0.
\end{equation}
In what follows we first analyze the effects of the nonextensive
distribution on the thermal conductivity $K_q$. It will be assumed
that the heat flux is macroscopically governed by the classical
Fourier law
\begin{equation}\label{eq26}
{\bf q}_e = - K_q{\bf \nabla} T,
\end{equation}
while,  microscopically, it is expressed as the usual average
value of the perturbed kinetic energy flow\cite{huang,KT}
\begin{equation}\label{eq17}
{\bf q}_e = \frac{1}{2}m\int v^2{\bf v}g d^3 v.
\end{equation}
As one may check, in the absence of external forces, the BGK model
leads to
\begin{equation}\label{g}
g=-\tau{\bf v}\cdot{\partial f_0\over\partial{\bf r}}.
\end{equation}
Substituting (\ref{g}) into (\ref{eq17}), and rewriting the
resulting expression in spherical coordinates $(v,\theta,\phi)$,
one obtains
\begin{equation}\label{eq19}
{\bf q}_e = - \frac{1}{2}m\int_0^\infty v^4\tau dv \int_0^\pi
\sin\theta d\theta \int_0^{2\pi}{\bf v}\left[{\bf
v}\cdot\frac{\partial {\it f_0}}{\partial{\bf r}}\right] d\phi.
\end{equation}
Before proceed further, we note that the orthogonality relation
\begin{equation}\label{eq20}
\int_0^\pi \int_0^{2\pi}v_iv_j \sin\theta d\theta d\phi =
\frac{4\pi}{3}v^2\delta_{ij}
\end{equation}
where $i,j \equiv x, y, z$, may be used to show that the angular
integral can be written as
\begin{equation}\label{eq21}
\int_0^\pi \sin\theta d\theta \int_0^{2\pi}{\bf v}\left[{\bf
v}\cdot\frac{\partial {\it f_0}}{\partial{\bf r}}\right] d\phi =
\frac{4\pi}{3}v^2 \frac{\partial {\it f_0}}{\partial{\bf r}},
\end{equation}
with equation (\ref{eq19}) simplifying to
\begin{equation}\label{eq22}
{\bf q}_e = - \frac{2\pi}{3}m\int_0^\infty v^6\tau \frac{\partial
{\it f_0}}{\partial{\bf r}} dv.
\end{equation}
As remarked earlier, the concentration $n$ and the temperature $T$
in the local $q$-distribution are spatially dependent while the
pressure remains constant. Thus, as a consequence of the extended
virial theorem\cite{PL}, one may write $n(r)T(r) = constant$, and
combining that with the spatial gradient of the unperturbed
distribution, the heat flow can be rewritten as
\begin{eqnarray*}
{\bf q}_e = - \frac{2\pi}{3}m\tau B_q\int_0^\infty v^6\{[1-(1-q)
\frac{m{\bf v}^2}{2k_BT}]^{q/(1-q)}\frac{mv^2}{2k_BT}
\end{eqnarray*}
\begin{equation}\label{we}
 - \frac{5}{2}\frac{1}{T} [1-(1-q)
\frac{m{\bf v}^2}{2k_BT}]^{1/(1-q)}\}\frac{\partial {\it
T}}{\partial{\bf r}}dv.
\end{equation}
Now, comparing with the Fourier law (\ref{eq26}), we obtain an
integral expression for the nonextensive thermal conductivity
\begin{eqnarray*}
K_q = \frac{\pi\tau m^2}{3k_BT}B_q\int_0^\infty v^8[1-(1-q)
\frac{m{\bf v}^2}{2k_BT}]^{q/(1-q)}dv
\end{eqnarray*}
\begin{equation}\label{eq27}
 - \frac{5\pi\tau m}{3T}
B_q\int_0^\infty v^6[1-(1-q)\frac{m{\bf v}^2}{2k_BT}]^{1/(1-q)}dv.
\end{equation}

The above integrals can easily be evaluated. We find
\begin{eqnarray*}
\int_0^\infty v^8[1-(1-q)\frac{m{\bf v}^2}{2k_BT}]^{q/(1-q)}dv=
\end{eqnarray*}
\begin{equation}\label{eq28}
\frac{1}{2}\left(\frac{2k_BT}{m}\right)^{9/2}\frac{105\pi^{1/2}}
{(9-7q)(7-5q)(5-3q)(3-q)}\left(\frac{1}{1-q}\right)^{1/2}
\frac{\Gamma({1\over 1-q})}{\Gamma(\frac{1}{2}+{1\over 1-q})}
\end{equation}
and,
\begin{eqnarray*}
\int_0^\infty v^6[1-(1-q)\frac{m{\bf v}^2}{2k_BT}]^{1/(1-q)}dv=
\end{eqnarray*}
\begin{equation}\label{eq29}
\frac{1}{2}\left(\frac{2k_BT}{m}\right)^{7/2}\frac{30\pi^{1/2}}
{(9-7q)(7-5q)(5-3q)(3-q)}\left(\frac{1}{1-q}\right)^{1/2}
\frac{\Gamma({1\over 1-q})}{\Gamma(\frac{1}{2}+{1\over 1-q})}.
\end{equation}
As one may check, inserting these integrals and the normalization
constant $B_q$ into equation (\ref{eq27}), we obtain
\begin{equation}\label{eq30}
K_q = \frac{5}{2}\frac{n\tau
k_B^2T}{m}\left[\frac{4}{(9-7q)(7-5q)}\right].
\end{equation}
The nonextensive parameter $q$ in the above expressions are
restricted to positive values with $q\neq9/7$,$7/5$. Note also
that expression (\ref{eq30}) reduces to standard value in the
extensive limiting case\cite{KT,reif}
\begin{equation}
K_1= \frac{5}{2}\frac{n\tau k_B^2T}{m}.
\end{equation}
It thus follows that the ratio between the nonextensive thermal
conductivity and the extensive value, $K_q/K_1$, is dependent on
the Tsallis' thermostatistics through parameter $q$ (see Fig. 1).
Note that for values of $q$ smaller than unity, the corresponding
coefficient $K_q$ can be much smaller than the standard result
(nearly $K_1/3$ for $q \sim 0.7$), while for $q>1$ it may
increases without limit.

Let us now consider the nonextensive effects on the viscosity
coefficient. In this case, the particles of the gas do not move
with the same velocity. In a simplified treatment, the particles
have constant mean velocity $u_x$ in the $x$ direction with the
magnitude of $u_x$ depending only on $z$, that is, $u_x =
{u_x}(z)$. Thus, neglecting bulk viscosity, the Navier-Stokes
stress reduces to

\begin{equation}\label{eq39}
P_{zx} =  - {\eta_q}\frac{\partial{u_x(z)}}{\partial{z}}.
\end{equation}
A complementary hypothesis is that the collisional state tend to
produce a local equilibrium distribution relative to the moving
gas with mean velocity $u_x$. Under such conditions, the
stationary $q$-distribution becomes
\begin{equation}\label{eq31}
f_0(v_x - u_x(z), v_y, v_z) = B_q [1-(1-q)\frac{m{\bf
U}^2}{2k_BT}]^{1/(1-q)},
\end{equation}
where $U_x = v_x - u_x(z)$, $U_y = v_y$, $U_z = v_z$. Since the
resulting  distribution $f$ is stationary but also depends on $z$
(the direction of the gradient velocity), the BGK equation
(\ref{eq1}) now reads
\begin{equation}\label{eq32}
v_z\frac{\partial{\it f}}{\partial{z}} = -\frac{\it{f-f_0}}{\tau}.
\end{equation}
The perturbed distribution function obeys
\begin{equation}\label{eq33}
g = -\tau v_z \frac{\partial{\it f_0}} {\partial{z}},
\end{equation}
so that
\begin{equation}\label{eq34}
f = f_0 + \tau v_z\frac{\partial{\it f_0}}
{\partial{U_x}}\frac{\partial{u_x(z)}}{\partial{z}}.
\end{equation}
In order to calculate the component $P_{zx}$ of the stress tensor,
we consider its kinetic definition\cite{huang}
\begin{equation}
P_{ij}=\int d{\bf U} U_i U_j f.
\end{equation}
Following standard lines we see that (for details see Ref.
\cite{reif})
\begin{equation}\label{eq38}
P_{zx} = m\int d^3U \tau \frac{\partial{\it f_0}}
{\partial{U_x}}U_z^2U_x\frac{\partial{u_x(z)}}{\partial{z}},
\end{equation}
and comparing this result with the macroscopic equation
(\ref{eq39}) we find
\begin{equation}\label{eq40}
{\eta_q} = - m\tau\int d^3U \frac{\partial{\it f_0}}
{\partial{U_x}}U_z^2U_x,
\end{equation}
where
\begin{equation}\label{eq41}
\frac{\partial{\it f_0}}{\partial{U_x}} = - \frac{B_q}{k_BT}
[1-(1-q)\frac{mU^2}{2k_BT}]^{q/(1-q)}mU_x
\end{equation}
Replacing (\ref{eq41}) into (\ref{eq40}) it follows that
\begin{equation}\label{eq42}
{\eta_q} = m\tau \frac{B_q}{k_BT}\int\int \int dU_xdU_ydU_z
[1-(1-q)\frac{m}{2k_BT}(U_x^2 + U_y^2 +
U_z^2)]^{q/(1-q)}U_z^2U_x^2,
\end{equation}
and performing the elementary integrals
\begin{equation}\label{eq43}
{\eta_q} = \frac{m^2\tau}{k_BT} B_q
\left(\frac{2k_BT}{m}\right)^{7/2}
\frac{2\pi^{3/2}}{(7-5q)(5-3q)(3-q)}\left(\frac{1}{1-q}\right)^{1/2}
\frac{\Gamma({1\over 1-q})}{\Gamma(\frac{1}{2}+{1\over 1-q})}.
\end{equation}
Finally, substituting the expression of $B_q$ and cancelling out
the common factors, the viscosity coefficient assumes the rather
simple form
\begin{equation}\label{eq44}
{\eta_q} = \frac{2}{(7-5q)} n\tau k_BT.
\end{equation}
This expression is valid only for $q\neq 7/5$, and as expected, in
the limit $q\rightarrow 1$ it reduces to the standard extensive
result\cite{KT}
\begin{equation}
\eta_1 = n\tau k_BT.
\end{equation}

In Figure 2 we show the ratio $\eta_q/\eta_1$ as a function of the
nonextensive parameter. This plot is similar to what happens with
the thermal conductivity (see Fig.1), and as expected, only for
$q=1$ the standard Maxwellian result is recovered. Note also that
combining equations (\ref{eq30}) and (\ref{eq44}) we also obtain a
quite simple expression to the dimensionless ratio, $\Lambda$,
involving the transport coefficients, $K_q$ and $\eta_q$, namely
\begin{equation}\label{ratio}
\Lambda \equiv {mK_q \over k_B\eta_q} = {5\over 9-7q}.
\end{equation}
Such a quantity (sometimes called Eucken's ratio\cite{woods})
plays an important experimental role because the unknown collision
time appearing in both transport coefficients cancels out in this
ratio. In Figure 3, we plot $\Lambda$ as a function of $q$. Note
that $\Lambda$ does not depend on the temperature, and for $q=1$
it reduces to the extensive result $\Lambda = 2.5$\cite{huang}. As
widely known, this pure number lies experimentally within the
range $1.3 \leq \Lambda \leq 2.5 $\cite{reif,Euc13}. Therefore,
the allowed values to the nonextensive parameter is restricted on
the interval $0.74 \leq q \leq 1$, thereby showing that the
Maxwellian result is only marginally compatible with such
measurements.

It is worth mentioning that the departure from standard Maxwellian
prediction is usually taken as an indication that the energy of
the molecules must include another forms than mere kinetic energy
of translation, or equivalently, an anomalous specific
heat\cite{Euc13}. However, as we have seen, a possible alternative
explanation (at least for monatomic gases) is provided by the
existence of nonextensive effects associated to Tsallis'
thermostatistics.

For completeness, we now investigate the possible nonextensive
effects on the electric conductivity of a dilute neutral plasma.
As before, we consider the BGK approximation assuming that the
external electric field is uniform and points to the positive $z$
direction, that is, ${\bf E} = E{\hat{\bf z}}$. The relevant
macroscopic equation is now the Ohm's law
\begin{equation}\label{eq11}
{\bf j_z}=\sigma_q E,
\end{equation}
where $\sigma_q$ is the electric conductivity. Microscopically,
the electric current density is defined as the statistical
average\cite{huang,KT}
\begin{equation}\label{eq9}
{\bf j}_n = e\int d^3 v f{\bf v}_n,
\end{equation}
where the subscript $n$ means that the charge flow is directed
along the normal to the corresponding surface element. Note that
only the component {\it j$_z$} does not vanish since for an
homogeneous and isotropic medium the current {\bf j} is parallel
to the electric field. As one may check, from equations
(\ref{eq0}), (\ref{oi}) and (\ref{eq9}) it follows that
\begin{equation}\label{eq10}
{\bf j}_z =\frac{e^2\tau E B_q}{k_BT}\int d^3 v
[1-(1-q)\frac{m{\bf v}^2}{2k_BT}]^{q/(1-q)}v_z^2.
\end{equation}
Performing the integration and comparing to the Ohm's law the
resulting electric conductivity is
\begin{equation}\label{eq12}
\sigma_q= \frac{e^2\tau}{k_BT}
B_q\pi^{3/2}\left(\frac{2k_BT}{m}\right)^{5/2}
\frac{2}{(3-q)(5-3q)}\left(\frac{1}{1-q}\right)^{1/2}
\frac{\Gamma({1\over 1-q})}{\Gamma(\frac{1}{2}+{1\over 1-q})}.
\end{equation}
or still, inserting the value of $B_q$ (see Eq. (\ref{eq13}))
\begin{equation}\label{eq14}
\sigma_{q}= \frac{ne^2\tau}{m}.
\end{equation}
Therefore, within the BGK approximation, the above cancellation
results that the electric conductivity does not depend on the
nonextensive $q$-parameter.

\section{Conclusion}

In this paper we discussed some possible nonextensive effects on
the transport phenomena in gases and plasmas. It was shown that
the theoretical analysis based on the relaxation-time
approximation lead to simple analytic expressions to the heat
conduction and viscosity coefficients, while the electric
conductivity was not modified.

It is interesting to compare our results (and the overall
approach) with independent analyzes and some previous expressions
to nonextensive transport coefficients appearing in the
literature\cite{boghosian,potiguar}. We first notice that our main
results are quite different from that ones obtained by
Boghosian\cite{boghosian}. Probably, the basic reason comes from
the fact that in his paper it was advocated a different choice to
the BGK operator, namely, $C_q = -{\tau}^{-1}({f}^q - {f_0}^q)$.
Note that it depends nonlinearly on the nonextensive parameter and
should be compared with the standard BGK operator assumed in the
present paper (see our equation (\ref{eq1})). In the treatment
adopted here, the nonextensive effects appear only implicitly
through the power law equilibrium distribution. In addition,
Boghosian (and also the authors of Ref. \cite{potiguar}) assumed
that the pressure in the perturbed gas does not remain constant
thereby giving rise to an anomalous heat conduction ($T_q$ in
their notation). In this concern, our approach may be considered
more conservative. As a matter of fact, using the extended virial
theorem \cite{PL}, we have assumed constant pressure as usually
done in the BGK approximation so that the anomalous heat
conduction coefficient, $T_q$, is absent. Another crucial
difference among these works is that three kinds of averaging
methods has been considered. Boghosian\cite{boghosian} and
Potiguar et al.\cite{potiguar} employed unnormalized and
normalized $q$-expectations values, respectively, while we have
computed the averages in the usual manner. In this way, it is not
surprising that although obtaining coefficients $K_q$ and $\eta_q$
proportional to the temperature, the behavior of the dimensionless
ratio $\Lambda \propto K_q/\eta_q$ as a function of the $q$
parameter is quite different (compare our equation (\ref{ratio})
with equations (83) and (40) in the quoted papers). Since this
ratio can be measured in laboratory, hopefully, these different
approaches, as well as their basic underlying assumptions will be
experimentally confronted in the near future.

{\bf Acknowledgments:} The authors are grateful to Uriel Costa for
helpful discussions. This work was supported by Pronex/FINEP (No.
41.96.0908.00), FAPESP (00/06695-0), CNPq and CAPES (Brazilian
Research Agencies).

\begin{figure}
\vspace{.2in}
\centerline{\epsfig{figure=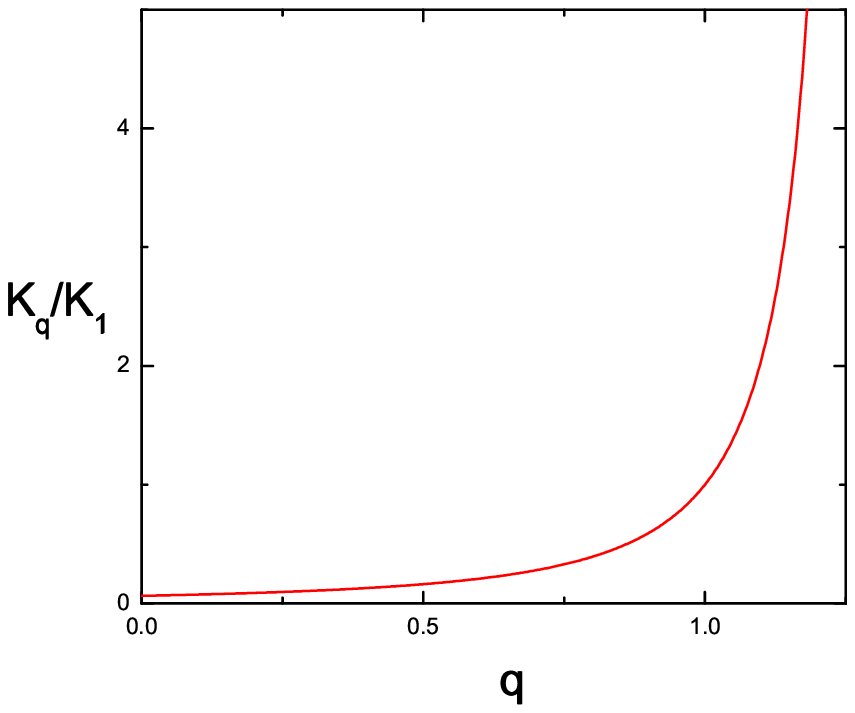,width=3.7truein,height=3.2truein}
\hskip 0.1in} \caption{Nonextensive thermal conductivity. We show
the dimensionless ratio $K_q/K_1$ as a function of the $q$
parameter. The normalizing quantity $K_1$ corresponds to the
Maxwellian expression obtained in the limit $q=1$. Note that for
$q\rightarrow 1.3$ the ratio $K_q/K_1$ increases with no limit.}
\label{gra1}
\end{figure}

\begin{figure}
\vspace{.2in}
\centerline{\epsfig{figure=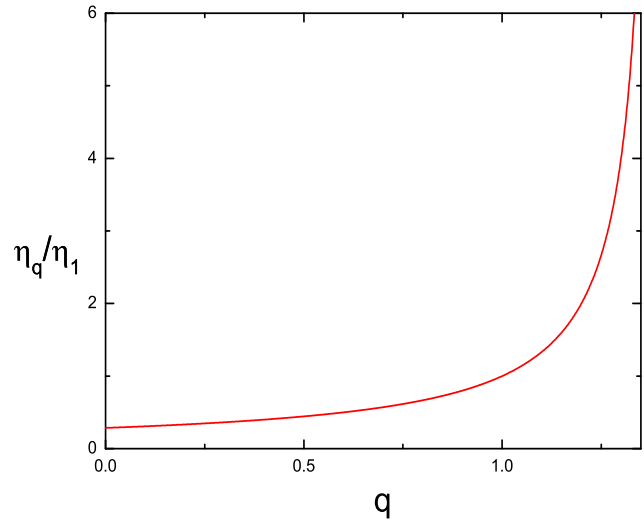,width=3.7truein,height=3.2truein}
\hskip 0.1in} \caption{Nonextensive viscosity coefficient. The
plot shows the dimensionless ratio $\eta_q/\eta_1$, where $\eta_1$
denotes the Maxwellian result, as a function of the nonextensive
parameter $q$. For $q\rightarrow 1.4$, this ratio goes to
infinity.}\label{gra2}
\end{figure}

\begin{figure}
\vspace{.2in}
\centerline{\epsfig{figure=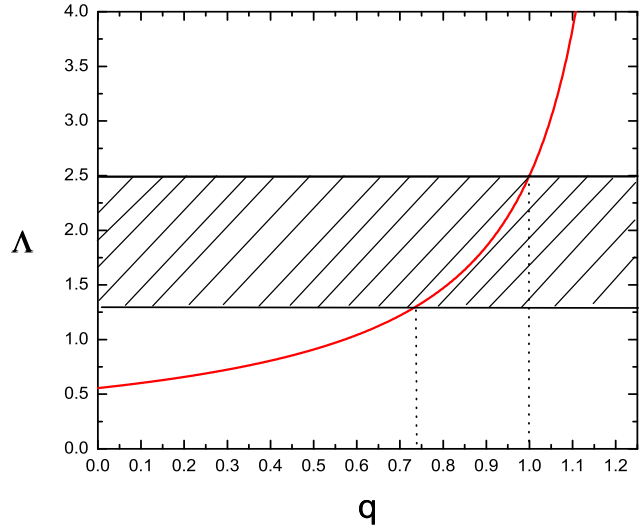,width=3.7truein,height=3.2truein}
\hskip 0.1in} \caption{The $\Lambda$ parameter. The experimental
dimensionless Eucken's ratio, $\Lambda = mK_q/{k_B\eta_q}$, is
shown as a function of the nonextensive parameter $q$. The
shadowed region shows the measured interval $1.3 \leq \Lambda \leq
2.5$, for which the $q$ parameter is restricted to $0.74 \leq q
\leq 1$. Note that the Maxwellian prediction is only marginally
compatible with such experiments.} \label{gra3}
\end{figure}

\end{document}